\documentclass[useAMS,usenatbib]{mn2e}

\usepackage{epsf}

\def\plotoneb#1#2{\centering \leavevmode
\epsfxsize=#2 \epsfbox{#1}}

\newcommand{\uxuma}{UX~UMa}
\newcommand{\oycar}{OY~Car}

\newcommand{\einstein}{{\sl Einstein\/}}
\newcommand{\rosat}{{\sl ROSAT\/}}
\newcommand{\asca}{{\sl ASCA\/}}
\newcommand{\xmm}{{\sl XMM-Newton\/}}

\newcommand{\xte}{{\sl RXTE\/}}
\newcommand{\nh}{N$_{\rm H}$}

\title[Two-Component X-ray Emission from UX UMa]{An \xmm\ observation of the nova-like variable UX UMa: spatially and spectrally resolved two-component X-ray emission}
\author[Pratt et al.]{G.W. Pratt$^{1,2}$\thanks{E-mail:
       gwp@mpe.mpg.de (GWP), mukai@milkyway.gsfc.nasa.gov (KM),
       bjmhassall@uclan.ac.uk (BJMH), T.Naylor@exeter.ac.uk (TN),
       jhw@astro.keele.ac.uk (JHW)}, K. Mukai$^{3}$\thanks{Also Universities
       Space Research Association}, B.J.M. Hassall$^4$,
       T. Naylor$^5$, and J.H. Wood$^6$ \\
   $^1$CEA/Saclay, Service d'Astrophysique, L'Orme des Merisiers,
           B\^at. 709, 91191 Gif-Sur-Yvette Cedex, France \\
   $^2$MPE Garching, Giessenbach{\ss}e, 85478 Garching, Germany \\
   $^3$Code 662, NASA/Goddard Space Flight Center, Greenbelt, MD 20771, USA \\
   $^4$Centre for Astrophysics, University of
           Central Lancashire, Preston PR1 2HE \\
   $^5$School of Physics, University of Exeter, Stocker Road,
           Exeter EX4 4QL \\
   $^6$Keele University, School of Chemistry and Physics, Keele,
           Staffordshire ST5 5BG }

\begin{document}

\date{Submitted 2003 October 22}

\pagerange{\pageref{firstpage}--\pageref{lastpage}} \pubyear{2004}

\maketitle

\label{firstpage}

\begin{abstract}

In the optical and ultraviolet regions of the electromagnetic
spectrum, UX~Ursae~Majoris is a deeply eclipsing cataclysmic variable.
However,  no soft X-ray eclipse was detected in \rosat\ observations.
We have obtained a 38 ksec \xmm\ observation to further constrain
the origin of the X-rays.  The combination of spectral and timing 
information allows us to identify two components in the X-ray emission of
the system.  The soft component, dominant below photon energies of 2 keV,
can be fitted with a multi-temperature plasma model and is uneclipsed.
The hard component, dominant above 3 keV, can be fitted with a $kT \sim 5$ keV
plasma model and appears to be deeply eclipsed. We suggest that the most
likely source of the hard X-ray emission in \uxuma, and other systems
in high mass transfer states, is the boundary layer.

\end{abstract}


\section{Introduction}

In non-magnetic cataclysmic variables (CVs), a white dwarf primary star is
accreting material, via an accretion disc, from a low mass quasi-main
sequence secondary star.  The disc instability theory predicts a limit
cycle behavior when the mass input rate from the secondary is low:
mass gradually accumulates in the disc at low optical brightness
(quiescence), then rapidly accretes onto the white dwarf during
outburst when the disc is hot and bright in the optical and in the UV
(see a recent review by \citealt{L2001}).  This is widely accepted
as the underlying cause of the dwarf nova phenomenon that is seen
in the majority of CVs.  In contrast, the mass input rate may be high
enough in some CVs for the accretion disc to achieve a steady state.
This is thought to be the explanation for nova-like systems, which resemble
dwarf novae in outburst at all times.

Simple models suggest that the boundary layer between the primary and
the accretion disc should be extremely luminous, emitting up to half
the accretion energy.  Since most of this is expected to be in X-rays
\citep{P1977}, X-ray observations have become an essential tool for
the study of accretion discs in CVs.  It was discovered early on that
dwarf novae in quiescence were moderately bright hard (2--10 keV) X-ray
sources.  In outburst, however, the hard component generally became
weaker, and sometimes bright soft ($<$1 keV) emission was observed.
Combining extensive \einstein\ X-ray data on CVs with theoretical
considerations, \cite{PR1985a,PR1985b} linked this to the optical depth
in the boundary layer.  For low accretion rates (appropriate for dwarf
novae in quiescence), the boundary layer is hot (kT$\sim$10 keV) and
optically thin, cooling via hard X-ray bremsstrahlung emission, while
for high accretion rates (dwarf novae in outburst and nova-like
systems), it becomes optical thick, emitting a blackbody-like spectrum
with kT$\sim$10--50 eV.  Optically thin, outer regions of the boundary
layer can explain the residual hard X-ray emission. The recent X-ray,
EUV, and optical observations of SS~Cygni through an outburst
\citep{Wea2003} are a recent confirmation of this basic picture.

Timing observations of eclipsing systems are invaluable as a method for
determining the source of the X-ray emission.  Observations of high
inclination dwarf novae in quiescence have shown evidence for a compact
X-ray source that is eclipsed with the white dwarf
\citep{Wea1995a,vT1997,Mea1997,Pea1999a,Rea2001},
although the X-ray emission in \oycar\ may be from a polar region
rather than from an equatorial boundary layer \citep{WW2003}.
However, the soft X-ray emission from the dwarf nova \oycar\ in
superoutburst\footnote{For the purpose of this paper,
normal outbursts and superoutbursts can be considered together.}
was found to be uneclipsed \citep{Nea1988}.  Based on a comparison
with lower inclination (non-eclipsing) systems, they advocated a
two-component origin for the soft X-ray emission from dwarf novae
in outbursts: a compact region, presumably the boundary layer,
and an extended region, possibly due to the scattering of emission from 
the former
in an accretion disc corona or a wind.  \cite{Nea1988} argued that
the white dwarf photosphere and the boundary layer are hidden from
our view at all orbital phases in \oycar\ in superoutburst, due to
the thickened rim of the accretion disc.  Later observations of
\oycar\ in superoutbursts confirm the lack of eclipse and
strengthen the scattering interpretation \citep{Pea1999b,MR2001}.

\xmm\ provides several key advantages in the study of eclipsing
CVs, such as the high effective area of the EPIC instruments both
at low and high energies.  We have chosen the nova-like system,
\uxuma, as our target for this study, which avoids the logistical
complication of scheduling an observation at short notice during an
outburst of a dwarf nova.  \uxuma\ is the brightest eclipsing nova-like
system in the optical and in the X-rays, and has been well studied in
ultraviolet in recent years. From UV data, \cite{Bea1995} claimed the
detection of a white dwarf eclipse, which however has been brought into
question by \cite{Fea2003}.  A \rosat\ PSPC observation by \cite{Wea1995b} 
showed no soft X-ray eclipse.

\section{Observation}

\begin{table}
\caption{Log of XMM-Newton observation 0084190201 of UX~UMa.}
\label{obs}
\begin{tabular}{lrrl}
\hline
Instrument & Start & End & Data Mode \\
\hline
MOS 1\&2 & 01:07 & 14:33 & Thin Filter, Full Frame \\
pn       & 01:30 & 14:23 & Thin Filter, Full Frame \\
RGS 1\&2 & 01:06 & 14:40 & Spectro+Q \\
OM       & 01:37 & 14:10 & UV Grism 1 \\
\hline
\end{tabular}

\medskip
All times in TT on 2002 June 12.
\end{table}

\uxuma\ was observed with \xmm\ on 2002 June 12 for approximately 45~ks
(see Table\,\ref{obs} for details).  Periods of high background in the EPIC
data due to soft proton solar flares were removed using the method described
in \citet{PA2003}.  There is one large flare in the middle of the observation,
otherwise the data are relatively clean\footnote{ Unfortunately, this
flare causes us to lose one of the 3 eclipses covered by these data.}.
Useful exposure times after cleaning were 37895s, 38412s and 28736s for
MOS1, MOS2 and pn, respectively. 

For the MOS cameras, PATTERNs 0--12 were selected; we use only the
well-calibrated single events (PATTERN 0) in the pn analysis. In
addition, we select FLAG=0 to exclude events at CCD edges and around
bad pixels.  

EPIC source light curves were extracted in 1s bins from a circular
region of radius $30\arcsec$. The background light curves were
accumulated in 1s bins from an annular region centred on UX UMa, with
inner and outer radii of $1\arcmin.45$ and $5\arcmin.3$,
respectively. Other sources were identified by eye and removed before
the background products were accumulated. The EPIC spectra were
extracted in the same regions as described above.  Response and
effective area files were produced using the SAS tasks {\tt rmfgen}
and {\tt arfgen}.

For the RGS data, the standard pipeline-produced spectra and response
matrix files were used. The `fluxed' RGS1+RGS2 spectrum shown in
Figure\,\ref{rgsspec} was produced from these files using the SAS
task {\tt rgsfluxer}.

\section{Results}

\subsection{EPIC spectral characteristics}
\label{sec:globspec}

\begin{figure*}
\begin{minipage}{144 mm}
\plotoneb{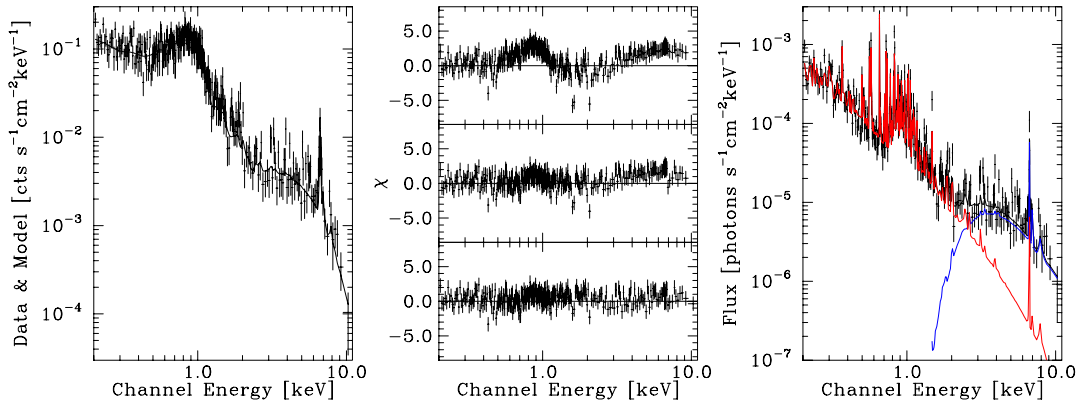}{144 mm}
\caption{The \xmm\ EPIC spectrum of \uxuma, as determined from
simultaneous fit to PN, MOS1, and MOS2 data: (Left) The observed PN spectrum
plotted with the best-fit model (see text) convolved with instrument
response. (Middle) The residuals, in the form of $\delta \chi$
for single temperature {\tt mekal} (top), {\tt cemekl} (middle), and
our best-fit model (bottom).  (Right) The two components of our best-fit
model are shown overplotted on deconvolved PN data. Units are flux units.}
\label{epicspec}
\end{minipage}
\end{figure*}

The EPIC spectra of \uxuma\ are shown in Figure\,\ref{epicspec}.
It is clear that the underlying emission is line-rich: the
Fe K$\alpha$ line is clearly visible, so is the ``bump'' around
1 keV indicative of the unresolved Fe L complex and other lines.
Since we wish to optimise the choice of spectral bands for the 
extraction of the light curves (Sect.~\ref{sec:lcrvs}), we will characterise 
the orbit-averaged spectrum of the system with the simplest plausible
model and use this as a constraint for the choice of light curve bands.

In view of the line-rich nature of the spectrum, we start with a
single-temperature {\tt mekal} model
\citep{Mea1985,Mea1986,Lea1995} for thermal emission from
collisionally excited plasma. This model fails to fit the data
($\chi_\nu^2$=3.6), and predicts more flux than observed around 2 keV
and less flux than observed in the 1 keV bump and at high energies.
Even a multi-temperature plasma does not fit well: with a {\tt cemekl}
\citep{Sea1996} model, in which the differential emission measure is a
power-law of plasma temperature, we obtain $\chi_\nu^2$=1.6 for 563
degrees of freedom (i.e., formally unacceptable).  Moreover, this fit
is achieved with best-fit abundances of 3.9 times Solar.  While there
is no compelling reason to believe \uxuma\ has exactly Solar
abundances, a factor of 3.9 seems too extreme to be believable.
Additionally, systematic residuals are evident (see middle panel of
Figure\,\ref{epicspec}) even with the overabundant {\tt cemekl} model.

\begin{table}
\caption{Results of single temperature {\tt mekal} model fits to the hard component in the EPIC spectra.}
\label{hardspec}
\begin{tabular}{rllll}
\hline
Energy Range & $\chi^2_\nu$/D.O.F. & \nh & kT & Norm. \\
\multicolumn{1}{c}{(keV)} & & \multicolumn{1}{c}{(cm$^{-2}$)} &
\multicolumn{1}{c}{(keV)} & \\
\hline
  5--10 & 1.25/ 54 & 1.0$\times 10^{22}$ & 5.6 & 4.2$\times 10^{-4}$ \\
  4--10 & 1.04/ 90 & 6.9$\times 10^{22}$ & 5.5 & 4.8$\times 10^{-4}$ \\
  3--10 & 1.06/125 & 5.6$\times 10^{22}$ & 5.5 & 4.4$\times 10^{-4}$ \\
  2--10 & 1.12/171 & 3.8$\times 10^{22}$ & 6.0 & 3.8$\times 10^{-4}$ \\
1.5--10 & 1.48/216 & 1.7$\times 10^{22}$ & 7.7 & 2.8$\times 10^{-4}$ \\
1.0--10 & 2.51/318 & 0.0$\times 10^{22}$ & 9.8 & 1.8$\times 10^{-4}$ \\
\hline
\end{tabular}
\end{table}

We have therefore investigated the soft and hard parts of the spectrum
separately.  The strength of the Fe K$\alpha$ line at 6.7 keV suggests
a strong contribution from thermal plasma with kT in the 1--10 keV
range (where Fe is mostly He-like). We thus experimented with the low
energy cutoff when fitting a single temperature {\tt mekal} model to
the hard part of the spectrum. When the low energy cutoff $E_{c}$ $\ga$ 2
keV (Table\,\ref{hardspec}), the fit becomes acceptable: including
energies below this threshold leads to a rapid deterioration in the
goodness of the fit, and a drastic change in the parameter
values. With $E_{c}$=3 keV, the best fit values (90\% confidence level
errors) are \nh=$(5.6\pm 1.7)\times 10^{22}$ cm$^{-2}$, kT=$(5.5\pm
1.1)$ keV, with a normalization of $4.4^{+1.1}_{-0.6} \times 10^{-4}$,
and a 3--10 keV flux of 3.2$\times 10^{-13}$ ergs\,s$^{-1}$cm$^{-2}$
(Solar abundances assumed). Note the high value of \nh, which would
indicate little contribution of the hard component below $\sim$2
keV. The temperature is constrained by the high energy cut-off and by
the dominance of the 6.7 keV line.  

A single-temperature, solar-abundance {\tt mekal} model fails to
fit the spectrum of \uxuma\ in the soft (0.2--2 keV) range ($\chi_\nu^2$=3.5
for 394 degrees of freedom).  Although it does much better ($\chi_\nu^2$=1.5)
when abundances are allowed to vary, the required abundances are 0.08 Solar,
a value that would make fitting of the hard range extremely problematic.
A better solution might be to use multiple-temperature plasma models.  Here
we have adopted {\tt cemekl}, resulting in $\chi_\nu^2$=1.2 (393 degrees of
freedom) with \nh=$7.8^{+8.4}_{-4.9}\times 10^{19}$ cm$^{-2}$, power law index
(for differential emission measure distribution)
$\alpha = 0.44^{+0.11}_{-0.16}$, kT$_{\rm max}$=$5.1^{+3.2}_{-1.2}$ keV,
a normalization of $1.8^{+0.2}_{-0.4} \times 10^{-4}$, with
a 0.2--2.0 keV flux of 2.3$\times 10^{-13}$ ergs\,s$^{-1}$cm$^{-2}$
(Solar abundances assumed).

\begin{table*}
\begin{minipage}{125mm}
\caption{Results of EPIC spectral fits in the 0.2--10 keV band.}
\label{combspec}
\begin{tabular}{lcccccc}
\hline
Comment & \nh & $\alpha$ & kT/kT$_{\rm max}$ &
        \multicolumn{3}{c}{Flux (10$^{-13}$ ergs\,s$^{-1}$cm$^{-2}$)} \\
	& (cm$^{-2}$) & & (keV) & 0.2--2 keV & 3--10 keV & 0.2-10 keV \\
\hline
Soft & $7.8^{+8.4}_{-4.9}\times 10^{19}$ & $0.44^{+0.11}_{-0.16}$ &
	$5.1^{+3.2}_{-1.2}$ & 2.28 & 0.37 & 2.87 \\
Hard & $8.1^{+1.3}_{-1.2}\times 10^{22}$ & & $5.7^{+1.3}_{-1.2}$ &
	0.0 & 2.85 & 3.05 \\
(Unabsorbed) & & & & 3.36 & 3.76 & 8.23 \\
\hline
\end{tabular}
\end{minipage}
\end{table*}

We show the result of the combined fit of these two models over the
0.2--10 keV range in Figure\,\ref{epicspec} and summarize the
parameters in Table\,\ref{combspec}.  We obtain $\chi_\nu^2$=1.14 for
568 degrees of freedom. The best-fit model to the soft part of the spectrum 
makes a small contribution to the 3--10 keV band and hence the parameters 
of the model fit to the hard part of the spectrum are somewhat modified 
from the results shown in Table\,\ref{hardspec}. 

\subsection{RGS spectral characteristics}

\begin{figure}
\plotoneb{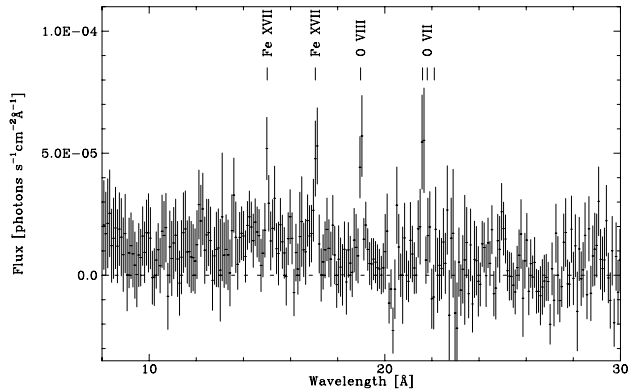}{84 mm}
\caption{The \xmm\ RGS spectrum of \uxuma, plotted against wavelength,
in the 6--30\AA\ range.  The four highest peaks are found at wavelengths
of two strong lines of Fe XVII, O VIII Ly$\alpha$, and the resonant
line of O VII (wavelengths of the intercombination and forbidden lines
are also indicated).}
\label{rgsspec}
\end{figure}

We present the combined RGS spectrum of \uxuma\ in
Figure\,\ref{rgsspec}.  Continuum is weakly detected at best,
while we detect 4 lines with some confidence.  We have fitted the
first-order spectra from both RGS 1 \& 2 simultaneously, after
rebinning the raw channels into groups of 4.
We have used a power-law continuum model with Gaussians to
obtain a rough characterisation of the lines (Table\,\ref{rgslines}).
Due to concern over the low signal-to-noise ratio in the continuum,
and the large contribution of the background to the total counts, we have
experimented with several different rebinning schemes. We can confirm
that the best fit values remain within the formal error ranges quoted
in Table\,\ref{rgslines}.
These 4 lines can be identified with Fe XVII 15.01 and 17.10 \AA,
(although the Fe XVII line at 17.05 \AA\ cannot be excluded as the
identification of the latter), O VIII Ly$\alpha$ (18.97 \AA), and
the resonance line of the O VII triplet (21.60 \AA).  The intercombination
(21.80 \AA) and forbidden (22.10 \AA) lines of O VII appear to be
significantly weaker.  Line widths are not significantly detected,
but we can establish weak upper limits (full-width at half-maximum, FMHM,
of 2000 -- 3000 km\,s$^{-1}$).

\begin{table}
\caption{Lines detected in the RGS spectra.}
\label{rgslines}
\begin{tabular}{rccl}
\hline
Species & Lab. $\lambda$ & Obs. $\lambda$ & Flux \\
        & (\AA) & (\AA) & (10$^{-6}$ photons\,s$^{-1}$cm$^{-2}$) \\
\hline
Fe XVII & 15.01 & 15.01$\pm$0.03 & 8.0$^{+3.0}_{-2.9}$ \\
Fe XVII & 17.10 & 17.09$\pm$0.04 & 9.3$^{+3.6}_{-3.9}$ \\
O VIII  & 18.97 & 19.00$\pm$0.02 & 11.2$^{+3.6}_{-3.5}$ \\
O VII   & 21.60 & 21.60$\pm$0.04 & 10.6$^{+5.4}_{-4.7}$ \\
\hline
\end{tabular}
\end{table}

\subsection{Folded Light Curves}
\label{sec:lcrvs}

We have seen in Sect.~\ref{sec:globspec} how the entire 0.2--10. keV
EPIC spectrum can be adequately fitted with a two component
model.  Considerable uncertainties remain in the model parameters, and 
the low energy part of the spectrum can also be described by more 
complex models (e.g., partial covering or partial ionisation) but
this simple and plausible spectral fitting serves to motivate the 
choice of bandpasses for the light curve extraction.

\begin{figure}
\plotoneb{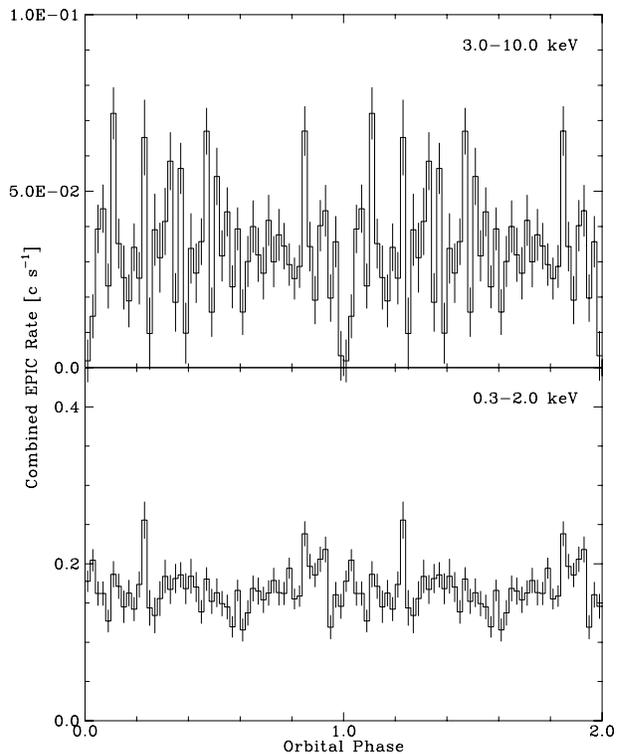}{84 mm}
\caption{The \xmm\ EPIC light curves of \uxuma, folded on the orbital
ephemeris of Baptista et al. (1995)
%\cite{Bea1995},
in 50 bins per cycle, in two energy bands.  The 0.3--3 keV light,
shown in the lower panel, is consistent with
no eclipse.  The 3--10 keV light curve (upper panel),
however, suggests the existence of a deep eclipse.}
\label{epicfold}
\end{figure}

We have thus extracted the light curves of \uxuma\ from
the three EPIC cameras in the energy bands 0.3--2.0 keV and 3.0--10 keV.
We then folded each of the 6 light curves on the orbital period, using the
ephemeris of \cite{Bea1995}\footnote{\cite{Fea2003} found it necessary to
shift their 2001 FUV light curves earlier in phase by 0.006 cycles.
We have not applied such a shift.}.  The folded light curves from the
three cameras were then averaged to produce a single folded light curve
per energy range.  These are shown in Figure\,\ref{epicfold}.
The light curves, particularly in the hard band, show considerable
variability, often in the form of flare-like events lasting several
hundered seconds.  These are not averaged out in the folded light
curve because the observation covered only a few cycles.

We confirm the \rosat\ result \citep{Wea1995b} that no eclipse
is seen in \uxuma\ in the soft X-rays ($<$2.0 keV).  However,
there does appear to be an eclipse in the hard ($>$3 keV) band,
at phase 0.0 as defined by optical/UV eclipse.  According to
\cite{Bea1995}, the white dwarf eclipse width (mid-ingress to
mid-egress) is 0.053$\pm$0.001 cycles, while totality (2nd to 3rd
contact) is a little over 0.04 cycles.  To investigate the
significance of a hard X-ray eclipse with an assumed duration
similar to that of the UV eclipse, 
we have re-folded the data in 25 bins per cycle, with one of the
bins centred on the nominal phase 0.0 using the ephemeris of \cite{Bea1995}.
The ``eclipse'' bin, (corresponding to the two lowest points in
Figure\,\ref{epicfold}) has a count rate of 2.5$\pm$1.7 $\times 10^{-3}$
cts\,s$^{-1}$ (the error from propagation of counting errors), while
the other 24 bins have 34.4$\pm$8.5 $\times 10^{-3}$ cts\,s$^{-1}$ (where 
the error is derived from the standard deviation of these bins, 
to take into account source variability - see e.g., \citealt{Wea1995b}).
We thus appear to have detected an eclipse at the 3.8 $\sigma$
level.

The hard X-ray eclipse thus appears real, but we are unable to place tight
constraints on the parameters.  We can reproduce the eclipse shape
with a wide range of parameters (e.g., a mid-eclipse phase of 0.0--0.01, 
a mid-eclipse
flux between 0--20\% of out-of-eclipse flux, an eclipse width of 0.04--0.06
cycles, and eclipse transitions of less than $\sim$0.01 cycles).

\section{Discussion}

The combination of the spectral and timing results strongly suggests
 that there are (at least)
two components in the X-ray emission from \uxuma.  The soft component
is unabsorbed and uneclipsed and therefore must originate from an extended
region.  The hard component, which we have newly discovered, is heavily 
absorbed and eclipsed.

The soft component in \uxuma\ probably has the same origin as that of
\oycar\ in superoutburst \citep{Nea1988,Pea1999b,MR2001}, and perhaps
that of DQ~Her \citep{Mea2003}.  That is, the likely origin is scattering
of centrally generated X-rays in an extended region.  The prominence 
of the resonance component of the O VII triplet in the RGS spectrum
(Figure\,\ref{rgsspec}) supports this interpretation. 
\cite{MR2001} make a strong case that the scattering medium is the 
accretion disc wind in the case of \oycar\ in
superoutburst.  This may well be the case in \uxuma, since the
system is known to have a strong wind \citep{Bea1995,Fea2003}.  Although
we have not been able to confirm this by resolving the X-ray lines,
our upper limits for their widths (FWHM of 2000 -- 3000 km\,s$^{-1}$)
are consistent with the FUV line widths (FWHM $\geq$1800 km\,s$^{-1}$;
Froning et al. 2003).

The inferred luminosity of the hard component is 
$1.26\times 10^{31}$ ergs\,s$^{-1}$ in the 0.2--10 keV band
for a distance of 345 pc \citep{Bea1995}.  This value is similar
to those found for the residual hard X-ray emission component
in non-eclipsing dwarf novae in outburst.  \cite{Bea2001} infer
a bolometric luminosity of $\sim 6 \times 10^{30}$ ergs\,s$^{-1}$
for Z~Cam in outburst from \asca\ data (where there is an additional
softer component, which we would not see in \uxuma\ due to the
absorber), while the hard X-ray component of SS Cyg in outburst
reaches a minimum at about $\sim 3 \times 10^{31}$ ergs\,s$^{-1}$
in 3--20 keV in the \xte\ data \citep{Wea2003}.  In all three systems,
the existing spectra can be fitted with a single temperature thermal
model with kT in the 5--10 keV range.  These similarities both in
luminosity and in temperature suggest that the hard X-ray component
that we have discovered in \uxuma\ is the same as the hard X-ray
emission seen in dwarf novae in outburst.

The detection of a hard X-ray eclipse allows us to rule out an
extended origin for this emission, such as was suggested (for SS Cyg
in outburst) by
\cite{Wea2003}. If the emission is coming from a compact region close to the
white dwarf, the most natural source at this position would be the
boundary layer. At first sight, this may seem an odd suggestion, since
the boundary layer is normally thought of as a source of soft X-rays,
and in an eclipsing system, is hidden behind the disc rim. However, UV
observations suggest that a disc rim can supply a column density of
$\sim 10^{22}$ cm$^{-2}$ in OY Car \cite{Hea1994}, and that the
absorber in \uxuma\ appears to result in deeper absorption in the UV
than that seen in \oycar\ \citep{Bea1998}. Furthermore, our spectral
fits suggest a column of this order of magnitude.  Pending more
detailed analysis, we suggest that the same material can be
responsible for both X-ray and UV absorptions. Crucially, such a
column is sufficient to extinguish a soft X-ray source, but will not
significantly affect X-rays above 5 keV. Thus these X-rays may well be
coming from the more tenuous parts of the boundary layer
\citep{PR1985a}. 

We require more sensitive hard X-ray observations of
\uxuma\ (such as a longer observation with \xmm) to use the eclipse
light curves to constrain the nature of the hard X-ray emission in
these systems.  At the same time, better quality spectra of the soft
component are required to study the kinematics of the soft X-ray
scattering region in \uxuma.

%\section*{Acknowledgments}

\end{document}